\RequirePackage{lineno}

\documentclass[aps,prl,twocolumn,superscriptaddress,showpacs,preprintnumbers,linenumbers,amsmath,amssymb]{revtex4}
 \usepackage{orcidlink}
\usepackage{graphicx} % Include figure files
\usepackage{dcolumn}  % Align table columns on decimal point
\usepackage{natbib}
\usepackage{hyperref} 
\usepackage[export]{adjustbox}

\graphicspath{{ps}}

\newcommand{\ECM}{\ensuremath{E_{\rm c.m.}}}

\input ./belle2sym.tex

\begin{document}
%\vspace*{-3\baselineskip}
%\resizebox{!}{3cm}{\includegraphics{belle.eps}}

%\linenumbers

\preprint{\vbox{ \hbox{   }
						\hbox{Belle Preprint: 2023-22 }
     					\hbox{KEK Preprint: 2023-44 }}}

                   %     \hbox{Version 7}
                    %    \hbox{Intended for {\it Physical Review Letter}}
                     %   \hbox{Authors: M. Nayak, S. Dey, O. Fogel, A. Soffer}
                      %  \hbox{Committee: Dmitri Liventsev (chair),} \hbox{Swagato Banerjee and Torben Ferber}
  		              % \hbox{hep-ex nnnn}

\title{ \quad\\[1.0cm] Search for a heavy neutral lepton that mixes predominantly with the tau neutrino}
%%% Paper:
%%% Journal:  Physical Review
%%% Contacts:
%%% Non-responding authors or those who said NO are commented out.
%%% ====================================================================
%%% Click the RELOAD button on your web browser to see the updated file.
%%% ====================================================================
%%% Use \input{pubXXX-orcid} to insert this material into your latex file.
\noaffiliation
 \author{M.~Nayak\,\orcidlink{0000-0002-2572-4692}} % 2371
 \author{S.~Dey\,\orcidlink{0000-0003-2997-3829}} % 5023
 \author{A.~Soffer\,\orcidlink{0000-0002-0749-2146}} % 2217
  \author{I.~Adachi\,\orcidlink{0000-0003-2287-0173}} % 2590
% \author{K.~Adamczyk\,\orcidlink{0000-0001-6208-0876}} % 2239
% \author{J.~K.~Ahn\,\orcidlink{0000-0002-5795-2243}} % 7423
  \author{H.~Aihara\,\orcidlink{0000-0002-1907-5964}} % 2223
  \author{S.~Al~Said\,\orcidlink{0000-0002-4895-3869}} % 6823
  \author{D.~M.~Asner\,\orcidlink{0000-0002-1586-5790}} % 4684
  \author{H.~Atmacan\,\orcidlink{0000-0003-2435-501X}} % 2538
% \author{V.~Aulchenko\,\orcidlink{0000-0002-5394-4406}} % 8183
% \author{T.~Aushev\,\orcidlink{0000-0002-6347-7055}} % 3747
  \author{R.~Ayad\,\orcidlink{0000-0003-3466-9290}} % 3766
% \author{T.~Aziz\,\orcidlink{-}} % 3523
  \author{V.~Babu\,\orcidlink{0000-0003-0419-6912}} % 5623
% \author{S.~Bahinipati\,\orcidlink{0000-0002-3744-5332}} % 2332
% \author{A.~M.~Bakich\,\orcidlink{0000-0001-8315-4854}} % 2115
% \author{Y.~Ban\,\orcidlink{-}} % 3503
% ...
% 2387
% \author{V.~Zhulanov\,\orcidlink{0000-0002-0306-9199}} % 4983
\collaboration{The Belle Collaboration}

%%%% >>>>> insert the authorlist here. BEFORE the abstract !!!!! <<<<<
%%%% >>>>> from the authorship confirmation web page
%%% Name the file author.tex and use \input{author} to insert into your latex file.
%\author{S. Dey, O. Fogel, M. Nayak, A. Soffer} \affiliation{Tel Aviv University, Tel Aviv, Israel}
%\collaboration{The Belle Collaboration}
%\noaffiliation
%% end author list

\begin{abstract}
We report a search for a heavy neutral lepton (HNL) that mixes predominantly with $\nu_\tau$. 
The search utilizes data collected with the Belle detector at the KEKB asymmetric energy $e^+ e^-$ collider.
The data sample was collected at and just below the center-of-mass energies of the $\Upsilon(4S)$ and $\Upsilon(5S)$ resonances and has an integrated luminosity of $915~\textrm{fb}^{-1}$, corresponding to $(836\pm 12)\times 10^6$ $e^+e^-\to\tau^+\tau^-$ events.
We search for production of the HNL (denoted $N$) in the decay $\tau^-\to \pi^- N$ followed by its decay via $N \to \mu^+\mu^- \nu_\tau$.
The search focuses on the parameter-space region in which the HNL is long-lived, so that the $\mu^+\mu^-$ originate from a common vertex that is significantly displaced from the collision point of the KEKB beams.
Consistent with the expected background yield, one event is observed in the data sample after application of all the event-selection criteria.
We report limits on the mixing parameter of the HNL with the $\tau$ neutrino as a function of the HNL mass.

\end{abstract}

\pacs{12.60.-i,13.35.-r,14.60.Pq}

\maketitle

%%%% >>>> keep the final version single-spaced
\tighten

{\renewcommand{\thefootnote}{\fnsymbol{footnote}}}
\setcounter{footnote}{0}

%%%%%%%%%%%%%%%%%%%%%%%%%%
%\section{Introduction}
%\label{sec:intro}

While the standard model (SM) has only left-handed, massless neutrinos, measurements of neutrino flavor oscillations~\cite{ParticleDataGroup:2022pth} demonstrate that neutrinos do have mass.
This suggests the possible existence of right-handed neutrino states, which carry no SM charges, with corresponding Yukawa and Majorana mass terms in the Lagrangian. 
A large ratio between the mass terms gives rise to a seesaw mechanism~\cite{Minkowski:1977sc,Yanagida:1979as,Glashow:1979nm,Gell-Mann:1979vob,Mohapatra:1979ia,Schechter:1980gr,Schechter:1981cv} that explains the small neutrino masses and predicts the existence of heavy mass eigenstates referred to as heavy neutral leptons (HNLs).
In particular, discovery of an HNL in the GeV mass scale can also explain the origin of the baryon asymmetry of the Universe via leptogenesis~\cite{Davidson:2008bu,Pilaftsis:2009pk,Shaposhnikov:2009zzb} and would motivate searches for an additional, keV-scale HNL that could be a dark matter candidate~\cite{Asaka:2005an,Asaka:2005pn,Boyarsky:2009ix,Boyarsky:2018tvu,Ghiglieri:2020ulj}.

Many experiments have searched for an HNL that mixes with the electron or muon neutrino and have set limits on the corresponding mixing coefficients $\Vesq$, $\Vmusq$~\cite{ATLAS:2022atq, CMS:2022fut,  CMS:2021dzb, LHCb:2020wxx, ATLAS:2019kpx,LHCb:2014osd,Belle:2013ytx,Belle:2022tfo,NuTeV:1999kej,DELPHI:1996qcc,CHARMII:1994jjr,NA3:1986ahv,CHARM:1985nku,WA66:1985mfx}. 
Fewer experiments have directly probed the mixing coefficient $\Vtausq$ of an HNL with the $\tau$ neutrino, which requires production and/or identification of a $\tau$ lepton.

The search reported here probes $\Vtausq$ in the ``Neutrino portal with tau-flavor dominance''scenario~\cite{Beacham:2019nyx}, in which the HNL mixes predominantly with the $\tau$ neutrino, and $\Vesq$ and $\Vmusq$ are negligible.
In the HNL mass range $300 < m_N < 1600~\mevcc$, this scenario has been directly studied with different methods by the DELPHI~\cite{DELPHI:1996qcc}, ArgoNeuT~\cite{ArgoNeuT:2021clc}, and \babar~\cite{BaBar:2022cqj} experiments.
In addition, reinterpretations~\cite{Boiarska:2021yho,Barouki:2022bkt} of other searches from the CHARM~\cite{CHARM:1983ayi,CHARM:1985nku} and WA66~\cite{WA66:1985mfx} Collaborations have resulted in tight limits on $\Vtausq$.

Among the direct searches, both the \babar~\cite{BaBar:2022cqj} search and the current search utilize the large sample of $e^+e^-\to\tau^+\tau^-$ events available at $e^+e^-$ $B$~factories. 
The \babar search employs a missing-energy signature.
The search reported here uses the displaced vertex (DV) formed by the decaying HNL, following the method of Ref.~\cite{Dib:2019tuj}, which is used here for the first time.

We search for production of the HNL in the decay $\tau^-\to N \pi^-$.
Being lighter than the $\tau$ lepton, the HNL undergoes the weak-neutral-current decay $N\to \nu_\tau \mu^+\mu^-$.
Thus, the signal signature is a pion that is prompt, i.e., originating from a point near the $e^+e^-$ interaction point (IP), and a $\mu^+\mu^-$ pair originating from a DV, i.e., a point significantly displaced from the IP.
The search is sensitive to a range of $m_N$ and $\Vtausq$ values for which the HNL is long-lived~\cite{Bondarenko:2018ptm} yet decays within the Belle tracking system.

%%%%%%%%%%%%%%%%%%%%%%%%%%
%\section{Data samples and detector}
%\label{sec:data}

The search is based on the full data sample recorded with the Belle~\cite{Belle:2000cnh, Belle:2012iwr} detector at the KEKB asymmetric-energy $e^+e^-$ collider~\cite{Kurokawa:2001nw, Abe:2013kxa}. 
The largest part of the data is a sample with an integrated luminosity of $703~\rm fb^{-1}$ collected at a center-of-mass (c.m.)\ energy $\ECM=10.58$~GeV, corresponding to the $\Upsilon(4S)$ resonance~\cite{ParticleDataGroup:2022pth}.
Smaller samples of 121, 89, and $2~\rm fb^{-1}$ were collected, respectively, at the $\Upsilon(5S)$ resonance and about $60$~MeV below the $\Upsilon(4S)$ and $\Upsilon(5S)$.
%$976~\rm fb^{-1}$~\cite{Belle:2012iwr} 
The total number of $e^+e^-\to \tau^+ \tau^-$ events is $N_{\tau\tau} = (836 \pm 12) \times 10^6$,
%~\cite{Banerjee:2007is}
where the uncertainty arises mostly from the luminosity measurement.

The Belle detector, described in detail in Refs.~\cite{Belle:2000cnh, Belle:2012iwr}, is a large-solid-angle magnetic spectrometer that consists of a silicon vertex detector, a 50-layer central drift chamber (CDC) that spans the radial region $8.3<r<86.3$~cm, an array of aerogel threshold Cherenkov counters (ACC),  % <- \v{C}erenkov 2007.08
%a barrel-like arrangement of 
time-of-flight scintillation counters (TOF), and an electromagnetic calorimeter (ECL) composed of CsI(Tl) crystals located inside a superconducting solenoid coil that provides a 1.5~T magnetic field.  
Finally, the KLM system consists of the iron flux-return located outside of the coil, instrumented to detect $K_L^0$ mesons and to identify muons.  

% {\bf SVD2+SVD1, up to experiment 37:}
%Two inner detector configurations were used. A 2.0 cm radius beampipe and a 3-layer silicon vertex detector was used for the first sample of $XXX \times 10^6 ~\tau^{+}\tau^-$ pairs, while a 1.5 cm radius beampipe, a 4-layer silicon detector and a small-cell inner drift chamber were used to record the remaining $XXX \times 10^6 ~\tau^{+}\tau^-$ pairs\cite{svd2}.  

%Electron identification is performed with the likelihood ratio $\mathcal{P}_e =\mathcal{L}_{e}/(\mathcal{L}_{e}+\mathcal{L}_{\bar e})$~\cite{Hanagaki:2001fz}, where $\mathcal{L}_{e}$ $(\mathcal{L}_{\bar e})$ is the likelihood value for the electron (non-electron) hypothesis. These likelihoods are determined from specific ionization ($dE/dx$) in the CDC, the number of photons in the ACC, the ECL cluster shape, and the ratio of energy deposited in the ECL to track track momentum. 
Electron identification is performed with a normalized likelihood $\mathcal{P}_e$, determined from specific ionization ($dE/dx$) in the CDC, the number of photons in the ACC, the ECL cluster shape, the ratio of energy deposited in the ECL to the track  momentum, and the track-ECL cluster position matching~\cite{Hanagaki:2001fz}. 
%Muon candidates are identified with $\mathcal{P}_\mu =\mathcal{L}_{\mu}/(\mathcal{L}_{\mu}+\mathcal{L}_{\pi} + \mathcal{L}_{K})$~\cite{Abashian:2002bd}, where the muon likelihood $\mathcal{L}_{\mu}$  is calculated using the hit profiles in the KLM relative to the CDC track extrapolation, and the pion and kaon likelihoods $\mathcal{L}_{\pi}$, $\mathcal{L}_{K}$ are determined from $dE/dx$, ACC, and TOF information. 
%We apply $\mathcal{P}_e > 0.9$ and $\mathcal{P}_\mu > 0.9$ to select the electron and muon candidates, respectively. 
Muon candidates are identified with a normalized likelihood $\mathcal{P}_\mu$ determined from $dE/dx$, ACC, TOF, and KLM information~\cite{Abashian:2002bd}. 

To optimize the event selection, model the background, and determine the signal efficiency, we use various Monte Carlo (MC) simulation samples. 
Background samples include the processes
$e^+e^- \to \tau^+\tau^-$, 
$e^+e^- \to q\bar q$ where $q$ is a $u$, $d$, $s$, or $c$ quark,
$e^+e^- \to B\bar B$,
$e^+e^- \to e^+e^-$,
$e^+e^- \to \mu^+\mu^-$, 
and $e^+e^- \to e^+e^- f\bar f$ (where $f$ is an electron, muon, or quark),
generated with the event generators
KKMC~\cite{Jadach:1999vf} with TAUOLA~\cite{Davidson:2010rw},
PYTHIA~\cite{Sjostrand:2006za},
EvtGen~\cite{Lange:2001uf} with PYTHIA, 
BHLUMI~\cite{Jadach:1991by}, 
KKMC, 
and AAFH~\cite{Berends:1986ig}, respectively.
Most background-MC samples correspond to integrated luminosities that are 5 times that of the data. 
While the MC samples for $e^+e^-\to e^+e^-$ and $e^+e^-\to e^+e^-f\bar f$ are smaller, the contribution of these events to the background is negligible.
Signal-MC samples of $e^+e^- \to \tau^+\tau^-$ events are generated with KKMC.
The generic decays of one $\tau$ lepton, referred to as the tag $\tau$, are simulated  with TAUOLA and PYTHIA according to the known branching fractions~\cite{ParticleDataGroup:2022pth}.
The decays of the signal $\tau$ and of the HNL are simulated with PYTHIA. 
Signal MC samples are generated for $m_N$ values between 300~and 1600~$\mevcc$ in 25-$\mevcc$-wide steps.
Depending on $m_N$, the HNL lifetime $\tau_N$ is simulated such that $c\tau_N$ is between $15$ and $30$~cm, chosen so that a large fraction of the simulated HNLs decay within the CDC.
Photon production via initial-state radiation is simulated in all but the $e^+e^-\to B\bar B$ events. In all MC samples, final-state radiation from charged particles is simulated with PHOTOS~\cite{Barberio:1993qi}, and the detector-response simulation is performed with GEANT3~\cite{Brun:1987ma}.

In the signal samples, the HNL decay is simulated with equal probability density in all regions of the $\mu^+\mu^-\nu_\tau$ phase space, ignoring spin correlations.
To correct for this, signal events are also generated with MadGraph5\_aMC@NLO~\cite{Alwall:2014hca} with the HeavyN model of Majorana neutrinos~\cite{HEAVYN} %(\url{https://feynrules.irmp.ucl.ac.be/wiki/HeavyN}) 
and an effective vertex of the form $\partial_\mu \pi \bar \tau \gamma^\mu (1-\gamma_5) N + H.c$ for the $\tau^-\to \pi^- N$ decay.
For each value of $m_N$, the PYTHIA-generated events are weighted according to the Dalitz-plot distribution of the MadGraph5\_aMC@NLO  events at the generator level.
All other kinematic-variable distributions are consistent among the PYTHIA and MadGraph5\_aMC@NLO events, and the impact on the efficiency of potential correlations between the angular distributions of the tag-$\tau$ and signal-$\tau$ decays are determined to be negligible.

%%%%%%%%%%%%%%%%%%%%%%%%%%
%\section{Event selection}
%\label{sec:cuts}

The event-selection criteria are decided by comparing distributions of simulated signal and background events.
Each event is divided into two hemispheres centered on the c.m.\  thrust axis~\cite{Brandt:1964sa, Farhi:1977sg}, which is determined from the tracks and photons in the event.
One hemisphere must contain exactly three tracks and the other exactly one track, taken as the signal-$\tau$ and tag-$\tau$ decays, respectively.
The sum of the track charges must be zero.
The prompt pion candidate and the tag-$\tau$ daughter-track candidate must have transverse and longitudinal impact parameters of $|dr| < 0.5$ and $|dz| < 2.0$~cm with respect to the IP.
The pion candidate must satisfy $\mathcal{P}_\mu < 0.01$ and $\mathcal{P}_e < 0.01$.
Pairs of photons, each with energy above $0.1$~GeV, are combined to form $\pi^0$ candidates if their invariant mass satisfies $0.115 < m(\gamma\gamma) < 0.152~\gevcc$, corresponding to about $\pm3.5$ standard deviations ($\sigma$) with respect to the $m(\gamma \gamma)$ resolution around the nominal $\pi^0$ mass~\cite{ParticleDataGroup:2022pth}.
Events containing a $\pi^0$ candidate are discarded.
The total lab-frame energy of photons with energies greater than 200~MeV that are not used to form $\pi^0$ candidates must be less than $1$~GeV, to reject hadronic background while allowing for initial-state radiation, machine background, and detector noise in signal events.

HNL candidates are reconstructed from a $\mu^+\mu^-$ pair, where the muons are  selected with the requirement $\mathcal{P}_\mu > 0.9$.
The muons are fitted to a common vertex that constitutes the DV.
Henceforth, the momenta of the muons are evaluated at the DV.
The $\chisq$ probability of the DV fit is required to be greater than $10^{-5}$.
The radial position of the DV must satisfy $\rdv > 15$~cm with respect to the CDC symmetry axis, to suppress background from prompt tracks, decays of $K_S^0$ mesons and hyperons, and particle interaction in dense detector material.
To further suppress background from prompt tracks while retaining high signal efficiency, each muon must satisfy $\rhit - \rdv > -2$~cm, where \rhit\ is the radial position of the smallest-radius CDC hit of the muon.
The c.m.-frame angle between the muons must satisfy $\cos\theta_{\mu^+\mu^-} > 0.5$ to suppress cosmic-ray background and ensure that the muons are consistent with originating from a boosted parent. 
To ensure consistency with the $\tau$ decay, we apply the invariant-mass requirement $m_{\pi {\rm DV}}=\sqrt{(p_\pi + p_{\rm DV})^2}/c < 1.776$~\gevcc, where $p_\pi$ and $p_{\rm DV}$ are the 4-momenta of the pion and of the $\mu^+\mu^-$, respectively

Events are rejected if they satisfy $420<\mpipi<520~\mevcc$, where $\mpipi$ is the dimuon mass calculated with the pion mass hypothesis for the two muons. 
This rejects $K_S^0\to\pi^+\pi^-$ decays in which the pions decay to or are misidentified as muons.
%While the $K_S^0\to\pi^+\pi^-$ mass resolution is only a few MeV, this conservative requirement is applied in order to also reject decays in which a pion emits a hard photon or its momentum is grossly mismeasured.

Despite the neutrino in the final state, the constraints of the signal decay enable reconstruction of the full kinematics of the signal-$\tau$ decay chain with a twofold ambiguity~\cite{Dib:2019tuj}, which arises from a quadratic equation in the magnitude $p_N$ of the HNL momentum,
%%%%%%%%%%%%%
\begin{equation}
    (B^2-1)p_N^2 + (2AB-D)p_N + A^2 -C = 0.
    \label{eq:quad}
\end{equation}
%%%%%%
Here we have defined (with all quantities in the laboratory frame)
%%%%%%%%%%%
\begin{eqnarray}
A&=&\left(m_\tau^2 + m_{\mu\mu}^2 - m_\pi^2\right)/2E, \nonumber\\
B&=&\left(q_{\mu\mu} + q_\pi\right)/E, \nonumber\\
C&=&\left(E_{\mu\mu}(m_\tau^2-m_\pi^2) - E_\pi m_{\mu\mu}^2\right)/E, \nonumber\\
D&=&2\left(E_{\mu\mu} q_\pi - E_\pi q_{\mu\mu}\right)/E,
\end{eqnarray}
%%%%
where 
$m_\tau$, $m_\pi$, and $m^2_{\mu\mu}$ are the masses of the $\tau$, $\pi^\pm$~\cite{ParticleDataGroup:2022pth}, and ${\mu\mu}$ system, 
$E=E_\pi+E_{\mu\mu}$,
$q_X=p_X \cos{\theta_{NX}}$,
$p_X$ and $E_X$ are the measured momentum and energy for $X=\pi$ or $X=\mu\mu$,
and $\theta_{NX}$ is the angle between $\hat p_X$ and $\hat p_N$. 
We take $\hat p_N$ to be the vector between the IP and the DV, neglecting the small $\tau$ flight distance.
The derivation of these expressions is explained in the Appendix~\cite{ref:supp}.
The discriminant $S$ of Eq.~(\ref{eq:quad}) tends to be larger for background than for signal.
We require it to satisfy $S<0.4~{\rm GeV}^2$, retaining more than 98\% of signal events.
Given a solution for $p_N$, the squared HNL mass is 
\begin{equation}
    m_N^2=(D + C p_N)/E^2.
\end{equation}
The two solutions for $m_N$ are referred to as $m_+$ and $m_-$, depending on the sign in front of the square root of the quadratic-equation solution. 
For events with $S<0$ we set $S=0$, in which case $m_+=m_-$.
%For signal events, either $m_+$ or $m_-$ is consistent with the true HNL mass within the experimental resolution.

Events that pass these selections are divided into two signal regions (SRs). 
The region SRH, defined by the requirement $\mpipi>520~\mevcc$, targets heavy HNLs. 
Light HNLs are targeted by the region SRL, which is defined by $\mpipi<420~\mevcc$.
Furthermore, events in the SRL are required to satisfy either $m_+ <900$ or $m_- < 600~\mevcc$. 

%%%%%%%%%%%%%%%%%%%%%%%%%%
%\section{Background}
%\label{sec:bgd}

In the background-MC sample, we find two background events in the SRH, both from $e^+e^-\to\tau^+\tau^-$. 
One event contains the decay $\tau^-\to\nu_\tau \pi^-K^+K^-$, and the DV is formed from the muons originating from the decays $\pi^- \to \mu^- \bar\nu_\mu$ and $K^+\to \mu^+\nu_\mu$.
The other event contains $\tau^-\to\nu_\tau \pi^-K^0_S$, with both pion daughters of the $K^0_S$ decaying to the muons that constitute the DV.
In the SRL the MC background yield is two $e^+e^-\to\tau^+\tau^-$ events and two $e^+e^-\to q\bar q$ events. 
In all four events, the DV is formed from a muon produced in a pion decay and either another such muon or a pion from a $K_L^0$ or $K_S^0$ decay.
%
%In MC, the final background yield is two $e^+e^-\to\tau^+\tau^-$ events in the SRH, and two $e^+e^-\to\tau^+\tau^-$ events and two $e^+e^-\to q\bar q$ events in the SRL. 
%More details on these events are given in the supplemental material~\cite{ref:supp}.
%
Given that the integrated luminosity of the data is $1/5$ that of the MC samples, the MC prediction for the background yield is $0.40\pm 0.28$ in the SRH and $0.80\pm 0.40$ in the SRL, where the uncertainties arise from the finite MC sample size.

In addition to the use of MC samples, we study the background from the event yields in control regions (CRs) labeled CRH and CRL, in correspondence to the signal regions SRH and SRL.
Each CR is defined by the same selection criteria as the corresponding SR, except that only one of the tracks forming the DV is identified as a muon with $\mathcal{P}_\mu > 0.9$, while the other must satisfy $\mathcal{P}_e < 0.1$ and $\mathcal{P}_\mu < 0.1$ to suppress leptons and enhance the pion contribution.
In the background-MC sample, the CRH has 337 $e^+e^-\to \tau^+\tau^-$ events.
The DVs in these events are formed from pions produced in $K_S^0\to \pi^+\pi^-$ decays (in 76\% of the events), $K_S^0\to \pi^+\pi^-$ followed by pion decay to a muon (13\%), prompt $\pi^\pm$ or $K^\pm$ mesons (2\%) or muons produced in their decays (4\%), $K_L^0$ decays (2\%) and hadronic interactions in material (2\%).
The CRH also contains 26 $e^+e^-\to q\bar q$ events, where the DV is formed from $K_S^0$ decay (38\%), $\Lambda\to p\pi^-$ (27\%), $\pi^\pm$ or $K^\pm$ decays (12\%), particles produced in hadronic interaction in detector material (12\%), $K_L^0$ decay (8\%), and prompt $\pi^\pm$ or $K^\pm$ (4\%).
The CRL has 101 $e^+e^-\to \tau^+\tau^-$, in which the DVs are formed from $K_L^0$ decay (57\%), $e^+e^-$ from photon conversion (26\%), prompt $\pi^\pm$ or $K^\pm$ (15\%) and $K_L^0$ decay (2\%).
The CRL also has 81 $e^+e^-\to q\bar q$ events, with DVs formed from $\Lambda\to p\pi^-$ (59\%), $K_L^0\to \pi\mu\nu$ (40\%), and $K_S^0$ decays (1\%).
%In MC, CRH has 337 $e^+e^-\to \tau^+\tau^-$ events and 26 $e^+e^-\to q\bar q$ events.
%The CRL has 101 $e^+e^-\to \tau^+\tau^-$ and 81 $e^+e^-\to q\bar q$ events.
%
%Details on the origin of the DV tracks in the CR MC events are given in the supplemental material~\cite{ref:supp}.

Several validation regions (VRs), titled VRH$\pi\pi$, VRL$\pi\pi$, VR$K_S$, VRHss, and VRLss, are used for data-MC consistency studies and systematic uncertainty estimation.
The VRH$\pi\pi$ and VRL$\pi\pi$ are defined by the same selection criteria as the SRH and SRL, respectively, except that both DV tracks must satisfy the nonlepton requirement $\mathcal{P}_e < 0.1$, $\mathcal{P}_\mu < 0.1$. 
Data-MC comparison in these regions helps validate the background estimated for DVs produced from pairs of hadrons. 
The VR$K_S$, defined identically to VRH$\pi\pi$  but with $480 < \mpipi < 515~\gevcc$, is used to check the overall level of background from $K_S$ decays. 
The VRHss and VRLss are defined by the same criteria as SRH and SRL, respectively, except that the electric charges of the two muons have the same sign, opposite the charge of the prompt pion.
These regions are used to validate the level of potential background from coincidental crossing of tracks. 
The event types that populate the VRs in the MC are listed in the Appendix~\cite{ref:supp}.
%To study background from DVs produced from hadrons we use events in the VRs VRH$\pi\pi$ and VRL$\pi\pi$.
%These satisfy the same criteria as those in the SRH and SRL, respectively, except that both DV tracks satisfy the non-lepton criterion $\mathcal{P}_e < 0.1$, $\mathcal{P}_\mu < 0.1$.
%The overall level of $K_S$ background is studied in VR$K_S$, defined identically to VRH$\pi\pi$  but with $480 < \mpipi < 515~\gevcc$.
%%Therefore, the only difference between VRH$K_S$ and VRL$K_S$ is that the latter includes the requirement $m_+ <900~\mevcc$ or $m_- < 0.6~\mevcc$.  
%
%Data-MC comparison of potential background from accidental crossing of tracks is performed with the VRs VRHss and VRLss. 
%These are defined by the same criteria as SRH and SRL, respectively, except that the electric charges of the two muons have the same sign.

%%%%%%%%%%%%%%%%%%%%%%%%%%
%\section{Results}
%\label{sec:results}

For each region, the event yields observed in the data, the corresponding MC prediction, their ratio, and the statistical consistency between them are shown in Table~\ref{tab:stat}.
Also shown are the postfit yields in the SRs and CRs for the case of no signal.
The data contain one event in the SRH, with $m_+ = m_- = 1.473~\gevcc$, and no events in the SRL.
To avoid potential experimenter bias, the data event yields in the SRs were unveiled only after finalizing all analysis procedures and systematic uncertainty estimations.
In the CRs and VRs, the number of data events exceed the MC expectation by between 2\% and 43\%, and the naive data-MC statistical consistency $N^\sigma_{\rm obs,bgd}=(N_{\rm obs} - N_{\rm bgd})/\sqrt{N_{\rm obs} + \sigma^2_{\rm bgd}}$ ranges between 0.8 and 4.7.
While decays of $\tau$ leptons are well simulated, MC-data differences may arise from the simulation of $q\bar q$ events, which is not as well tested for low-multiplicity events.
The largest of these differences is used to estimate an uncertainty on the background model, described later.

%%%%%%%%%%%%%
 \begin{table}[!htbp]
    \centering
    \begin{tabular}{|lccccc|}
        \hline\hline
        Region      & $N_{\rm obs}$ & $N_{\rm bgd}$              &  $\frac{N_{\rm obs}}{N_{\rm bgd}}$ & $N^\sigma_{\rm obs,bgd}$ & Postfit \\
        \hline
        SRH         &  1   & $0.40\pm 0.28$  & 2.5 & 2.1 & $0.59 \pm 0.31$ \\
        SRL         &  0   & $0.80\pm 0.40$    & 0 & $-2.0$ & $0.69 \pm 0.45$ \\
        CRH         & 95   & $73.6 \pm 3.8$  & 1.29 & 2.0 & $93 \pm 8$ \\
        CRL         & 43   & $37.2\pm 2.7$   & 1.16 & 0.8 & $41 \pm 6$ \\
        \hline
        VRH$\pi\pi$ & 273  & $191\pm 6$      & 1.43 & 4.7 & \\
        VRL$\pi\pi$ & 165  & $127\pm 6$      & 1.30 & 2.7 & \\
        VR$K_S$    & 7917 & $7728\pm 39$    & 1.02 & 2.0 & \\
%        VRL$K_S$    & 277  & $ 198\pm 6$     & 1.40 & 4.4 \\
        VRHss       & 0    & $0.40\pm 0.28$  & 0    & & \\
        VRLss       & 0    & 0               & 0    & & \\
        \hline\hline
    \end{tabular}
    \caption{The number $N_{\rm obs}$ of events observed in the data, the expected number $N_{\rm bgd}$ of background events based on the MC simulation, the ratio $N_{\rm obs} / N_{\rm bgd}$, and the naive statistical consistency $N^\sigma_{\rm obs,bgd}=(N_{\rm obs} - N_{\rm bgd})/\sqrt{N_{\rm obs} + \sigma^2_{\rm bgd}}$ for each of the signal, control, and validation regions.
    The last column shows the postfit event yields for the SRs and CRs, obtained by maximizing the likelihood function when the signal yield is fixed to 0.
}
    \label{tab:stat}
\end{table}
%%%%%%

\begin{figure}[htb]
\begin{center}   
\includegraphics[width=1.0
\linewidth]{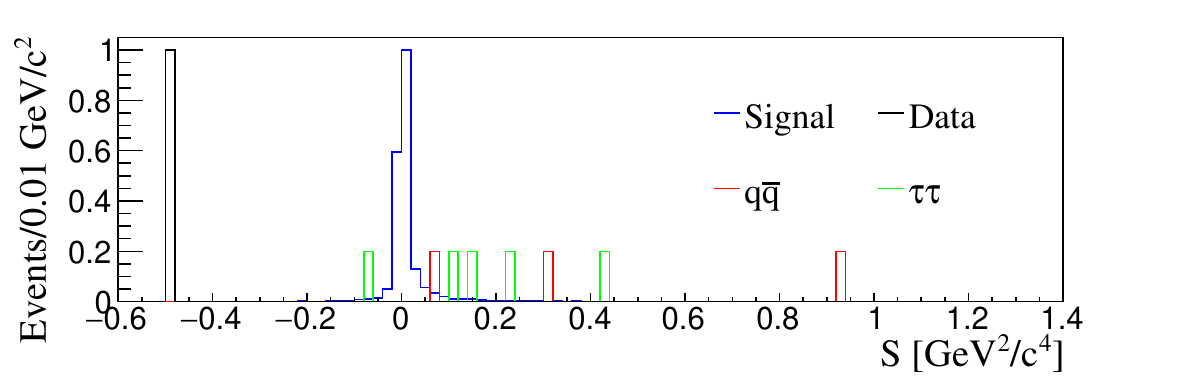}\\
(a)\\
\includegraphics[width=1.0
\linewidth]{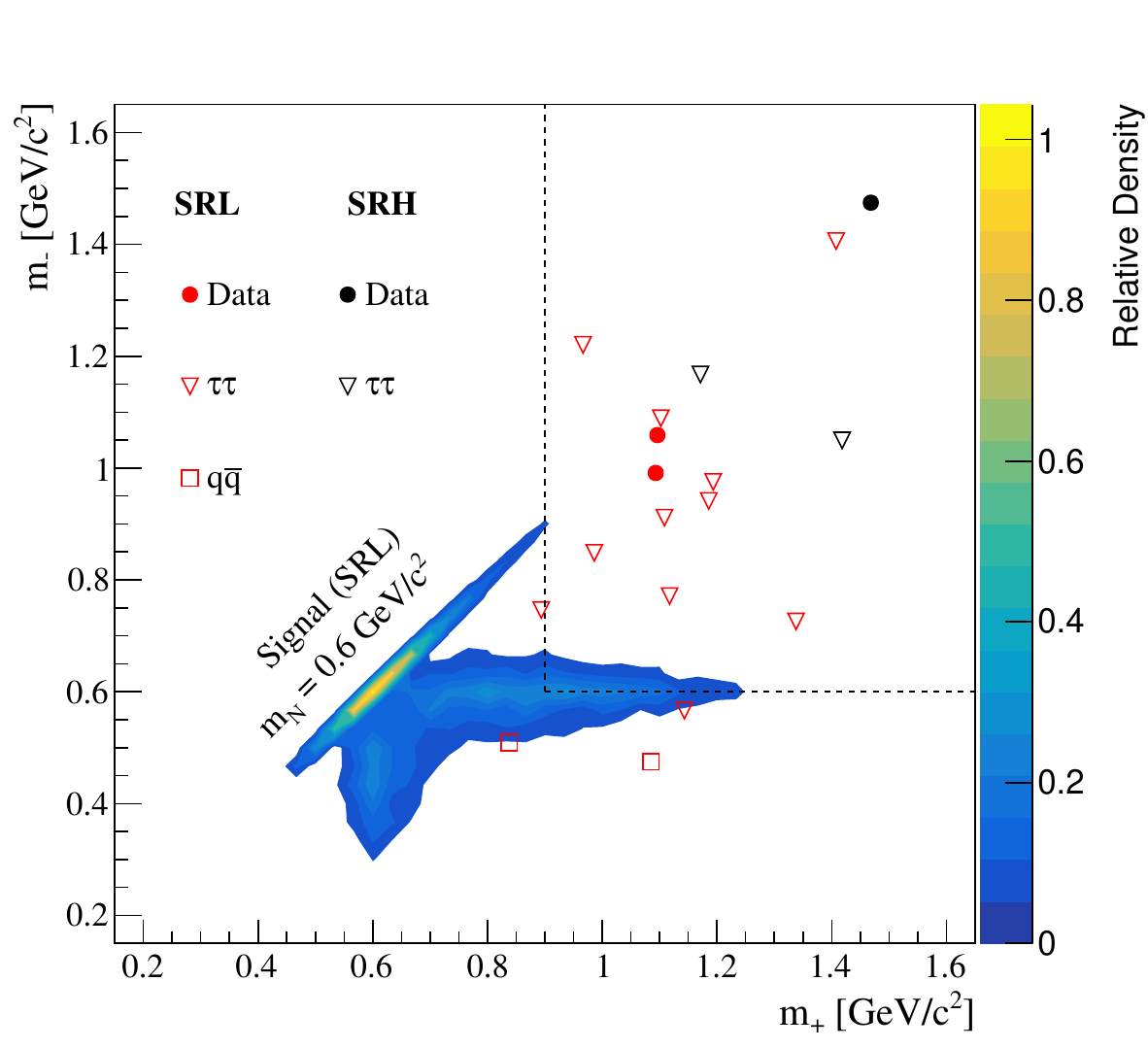}\\
(b)
\end{center}
\caption{(a) The $S$ values of the data and MC events after applying all SR requirements except $S<0.4$. 
The signal-MC distribution is arbitrarily normalized, and the background-MC distributions are normalized to the data luminosity.
(b) The $m_-$ vs. $m_+$ values for the data and for $e^+e^-\to\tau^+\tau^-$ and $e^+e^-\to q\bar q$ background-MC events (which has 5 times the data luminosity) in the SRH (black symbols) and SRL (red symbols). 
The region enclosed by the dashed lines is vetoed in the SRL.
The distribution of signal-MC events with $m_N=600~\mevcc$ in the SRL is also shown in colored contours.
}
\label{fig:mm} 
\end{figure}

The $S$ values of the data and background-MC events that pass all the selection criteria except $S < 0.4$ are shown in Fig.~\ref{fig:mm}(a) together with the signal-MC distribution.
The $m_\pm$ values after all selections except the $m_\pm$ requirement are shown in Fig.~\ref{fig:mm}(b).
The SRL distribution for signal MC events with $m_N = 600~\mevcc$ is also shown for comparison.
Signal events cluster around either $m_+\approx m_N$ or $m_-\approx m_N$, with events for which $S$ was set to $0$ having $m_- = m_+$.

%%%%%%%%%%%%%%%%%%%%%%%%

From the observed event yields $N_{\rm obs}$ in the SRs and CRs we compute 95\% confidence-level (CL) upper limits on $\Vtausq$ as a function of $m_N$ using the ${\rm CL_s}$ prescription implemented in pyhf~\cite{pyhf,pyhf_joss,Cranmer:1456844} with the likelihood function
\begin{equation}
L = \prod_R P \left(N^R_{\rm obs} | N^R_{\rm exp} \right)  
\prod_C G_C(p_C | p^0_C, \sigma_C).  
\label{eq:likelihood}
\end{equation}
%%%%%%%%%%%%%%%
Here the index $R$ runs through the regions SRH, SRL, CRH, and CRL; 
$P\left(N^R_{\rm obs} | N^R_{\rm exp} \right)$ is the Poisson probability for an observation of $N^R_{\rm obs}$ events in region $R$ given the expectation $N^R_{\rm exp}$;
the index $C$ runs through the nuisance parameters; and $G_C(p_C | p^0_C, \sigma_C)$ is the Gaussian distribution for nuisance parameter $p_C$ given the expectation $p_C^0$ and its uncertainty $\sigma_C$.

The expected event yield in region $R$ is $N^R_{\rm exp} = N_{\rm bgd}^R + N_{\rm sig}^R$, where $N_{\rm bgd}^R$ is the expected background yield shown in Table~\ref{tab:stat}, and $N_{\rm sig}^R$ is the expected signal yield, calculated as
\begin{align}
N^R_{\rm sig} = \sum_i 2 {\cal L}_i \sigma_i 
\BR(\tau^-\to \pi^- N) \BR(N\to \mu^+\mu^-\nu_\tau)  \epsilon^R .
\label{eq:Nsig}
\end{align}
%%%%%%%%%%%
Here the index $i$ indicates data samples with different values of $\ECM$;
${\cal L}_i$ is the integrated luminosity for sample $i$; 
$\sigma_i$ is the cross section for $e^+ e^-\to \tau^+\tau^-$, taken to be $0.919 \pm 0.003~{\rm nb}$ for $\ECM=\ECM(\Upsilon(4S))$ and scaled by $\ECM^2(\Upsilon(4S))/\ECM^2$ for other samples~\cite{Banerjee:2007is}; 
$\BR(\tau\to \pi N)$ and $\BR(N\to \mu\mu\nu)$ are the branching fractions of the specified decays~\cite{Bondarenko:2018ptm} listed in the Appendix~\cite{ref:supp}; 
and $\epsilon^R$ is the total efficiency for signal events to satisfy all the selection criteria of region $R$, determined from the signal-MC samples.
Since muons have a finite probability of failing the muon-identification criteria, signal events may populate the CRs.
The ratio $N^{\rm CR}_{\rm sig} / N^{\rm SR}_{\rm sig}$ between the signal yields in each CR and its corresponding SR ranges between $0.1$ and $2.1$, depending on the HNL mass. 
Nonetheless, the CRs are dominated by background events for all values of $\Vtausq$ for which the likelihood is not negligible.

The efficiency $\epsilon^R$ depends on $m_N$ and the HNL lifetime, which is taken from Ref.~\cite{Bondarenko:2018ptm} and listed in the Appendix~\cite{ref:supp} for each value of $m_N$ and $\Vtausq$.
While the signal-MC events are generated with a specific lifetime value $\tau_N^0$, we determine $\epsilon^R$ for any lifetime $\tau_N$, as follows.
First, we use each signal-MC sample to calculate the efficiency as a function of the radial and longitudinal position $(\rdv,\zdv)$ of the HNL decay, creating an efficiency map $\epsilon^R(m_N;\rdv,\zdv)$. 
The map bin size is 5~cm in $\zdv$ and 10~cm in $\rdv$.
Larger bins are used for the SRL (SRH) efficiency of heavy (light) HNLs, for which the sample size is limited.  %with bins of size $5~{\rm cm}\times 5~{\rm cm}$.
For each event in the signal-MC sample we randomly draw a set of decay times $t_i$ from the exponential distribution $\exp(-t_i/\tau_N)/\tau_N$. 
For each $t_i$ we calculate the event's HNL decay position and determine its efficiency from the efficiency map.
The individual event efficiencies are used to determine the total signal efficiency given the MadGraph5\_aMC@NLO  weights. 
The resulting signal efficiencies in the SRs are given in the Appendix.~\cite{ref:supp}. 
The calculation of $\epsilon^R$, $N_{\rm sig}^R$, and the upper limit is performed on a grid in $\Vtausq$-vs-$m_N$ space.
The grid uses a $25~\mevcc$ step in $m_N$, corresponding to the generated signal-MC samples. 
In $\Vtausq$ the grid is logarithmic, with a multiplicative step size of $10^{0.05}$.

All systematic uncertainties are handled with the nuisance parameters shown in Eq.~(\ref{eq:likelihood}).
The largest relative systematic uncertainty, 34\%, is assigned to the background yield expectations $N_{\rm bgd}^R$.
This value is chosen since it brings the event yields in the data and MC samples, $N_{\rm obs}^{\mathrm{VRH}\pi\pi}$ and $N_{\rm exp}^{\mathrm{VRH}\pi\pi}$, to within $1\sigma$ consistency in the VRH$\pi\pi$, which has the poorest data-MC consistency, as shown in Table~\ref{tab:stat}.
Following Ref.~\cite{ATLAS:2022atq}, we assign a 5\% uncertainty on the HNL branching fraction and decay modeling, arising mainly from the QCD corrections to the HNL hadronic decay width~\cite{Bondarenko:2018ptm,Davier:2005xq}.
The systematic uncertainty on the integrated luminosity is 1.4\%~\cite{Bevan:2014iga}, and that on the $e^+e^-\to \tau^+\tau^-$ cross section is 0.3\%~\cite{Banerjee:2007is}.
The uncertainty on the reconstruction of the two prompt tracks is 0.7\%~\cite{Belle:2020lfn}.
The relative statistical uncertainties on $\epsilon^R$ associated with the finite number of MC events, the finite number of generated decay times $t_i$, and the MadGraph5\_aMC@NLO  weights are also used as systematic uncertainties.
The uncertainty due to muon identification is 2\% per muon~\cite{Belle:2020lfn}.
The uncertainty on the efficiency for the remaining event-reconstruction steps, namely, online event selection and trigger, and tracking and vertexing of the HNL daughter tracks, is estimated to be 3.7\%, chosen so that it brings the data and MC yields to within $1\sigma$ consistency in the VR$K_S$.
We estimate an uncertainty of 1.3\% on the determination of the signal efficiency from the maximal difference with respect to the true efficiency at the generated lifetime.
%, and that on the trigger efficiency is 1.2\%.
The postfit values of the nuisance parameters are all well within $1\sigma$ of their expected values, and are shown in the Appendix~\cite{ref:supp}.
The largest pull is that of the background-prediction parameter $\mu_B$, $0.69\pm 0.43$.

\begin{figure}[htb]
\includegraphics[width=1\linewidth]{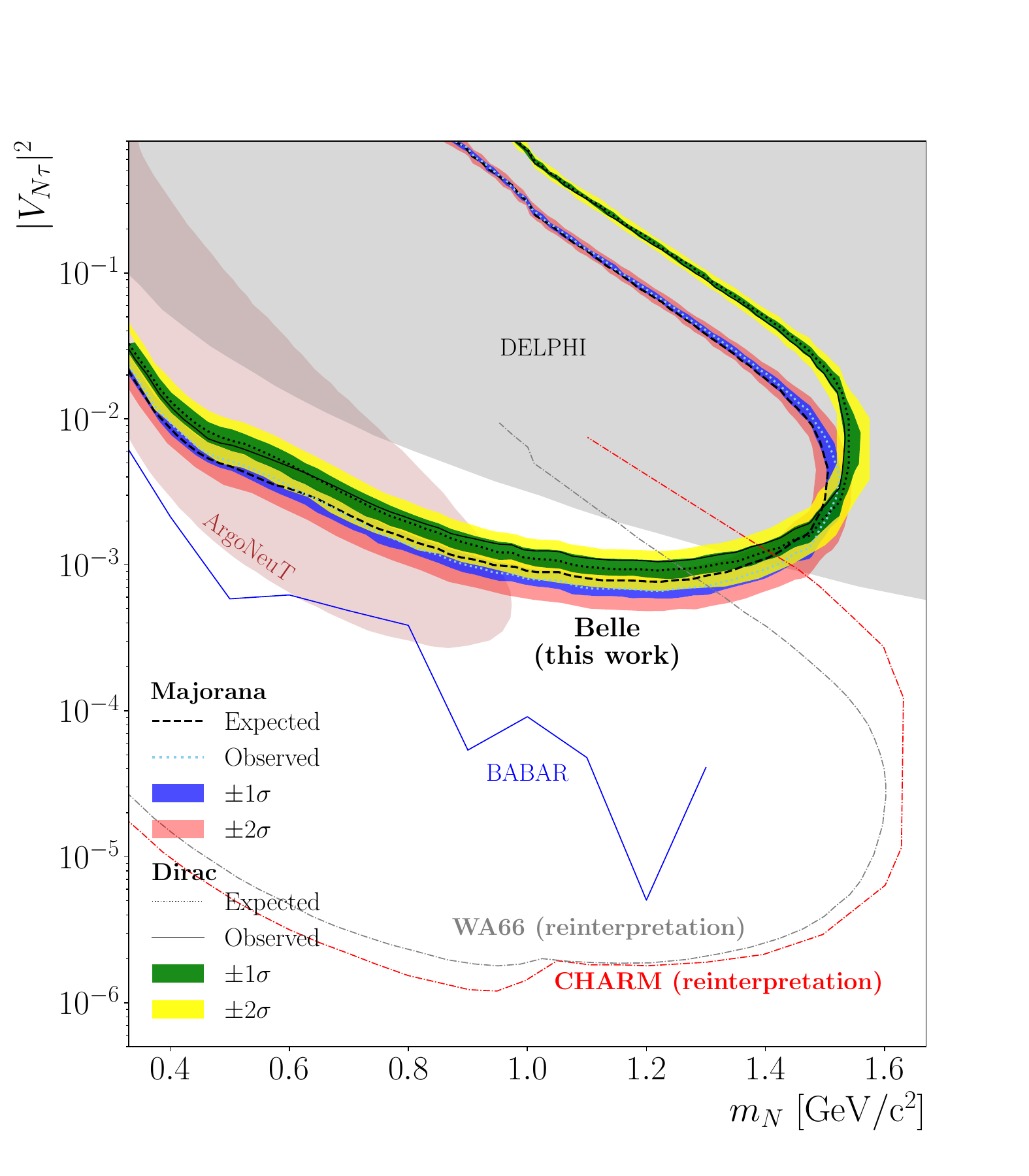}
\caption{The expected (dashed) and observed (solid) 95\% CL limits on $\Vtausq$ vs. $m_N$ for a Dirac or Majorana HNL. 
The green and yellow bands show the $1\sigma$ and $2\sigma$ bands for the expected limits for the Dirac case. The blue and pink bands show the same for the Majorana case.
Also shown are the limits from direct searches at DELPHI~\cite{DELPHI:1996qcc}, corrected for the unavailability of the charged-current decay for $m_N<m_\tau$~\cite{Helo:2011yg}, ArgoNeuT~\cite{ArgoNeuT:2021clc} (90\% CL), and the upper limit from BABAR~\cite{BaBar:2022cqj}.
The 90\% CL limits arising from reinterpretations~\cite{Boiarska:2021yho,Barouki:2022bkt} of other searches by CHARM~\cite{CHARM:1983ayi,CHARM:1985nku} and WA66~\cite{WA66:1985mfx} are shown as well.
}
\label{fig:limits_combined} 
\end{figure}

The resulting excluded region in $\Vtausq$-vs-$m_N$ space are shown in Fig.~\ref{fig:limits_combined}, separately for a Dirac and a Majorana HNL. 
For every point on the grid, the HNL lifetime in the Majorana case is half that of the Dirac HNL.

%This is the main result of the analysis.

%%%%%%%%%%%%%%%%%%%%%%%%%%
%\section{Conclusions}
%\label{sec:conclusions}

%\conclusion
In conclusion, we report a search for a heavy neutral lepton in the decay chain $\tau^-\to \pi^- N$, $N\to \mu^+\mu^-\nu_\tau$. 
The search method, used here for the first time, utilizes the displaced vertex originating from the long-lived HNL decay and the ability to reconstruct the HNL-candidate mass to suppress the background to the single-event level. 
It also allows for direct measurement of the HNL mass if a signal is observed.
In the signal regions targeting heavy and light HNLs we observe $1$ and $0$ events, respectively, in agreement with the background expectation.
We set limits on the mixing coefficient $\Vtausq$ of the HNL with the $\tau$ neutrino for HNL masses in the range $300 < m_N < 1600~\mevcc$.
We are grateful to N. Neill for help with HNL theoretical parameters and the MadGraph5\_aMC@NLO event generator.
This work, based on data collected using the Belle detector, which was
operated until June 2010, was supported by 
the Ministry of Education, Culture, Sports, Science, and
Technology (MEXT) of Japan, the Japan Society for the 
Promotion of Science (JSPS), and the Tau-Lepton Physics 
Research Center of Nagoya University; 
the Australian Research Council including Grants No.
DP210101900, % Urquijo
No. DP210102831, % Sevior
No. DE220100462, % Hsu
No. LE210100098, % Infrastructure
No. LE230100085; % Infrastructure
Austrian Federal Ministry of Education, Science and Research and Austrian Science Fund (FWF) [Grant DOI: 10.55776/P31361-N36];
National Key R\&D Program of China under Contract No.~2022YFA1601903,
National Natural Science Foundation of China and research Grants
No.~11575017,
No.~11761141009, 
No.~11705209, 
No.~11975076, 
No.~12135005, 
No.~12150004, 
No.~12161141008, 
and
No.~12175041, 
and Shandong Provincial Natural Science Foundation Project ZR2022JQ02;
the Czech Science Foundation Grant No. 22-18469S;
Horizon 2020 ERC Advanced Grant No.~884719 and ERC Starting Grant No.~947006 ``InterLeptons'' (European Union);
the Carl Zeiss Foundation, the Deutsche Forschungsgemeinschaft, the
Excellence Cluster Universe, and the VolkswagenStiftung;
the Department of Atomic Energy (Project Identification No. RTI 4002), the Department of Science and Technology of India,
and the UPES (India) SEED finding programs No. UPES/R\&D-SEED-INFRA/17052023/01 and No. UPES/R\&D-SOE/20062022/06; 
the Israel Science Foundation and the U.S.-Israel Binational Science Fund;
the Istituto Nazionale di Fisica Nucleare of Italy; 
National Research Foundation (NRF) of Korea Grants
No.~2016R1\-D1A1B\-02012900, No. 2018R1\-A2B\-3003643,
No. 2018R1\-A6A1A\-06024970, No. RS\-2022\-00197659,
No. 2019R1\-I1A3A\-01058933, No. 2021R1\-A6A1A\-03043957,
No. 2021R1\-F1A\-1060423, No. 2021R1\-F1A\-1064008, No. 2022R1\-A2C\-1003993;
Radiation Science Research Institute, Foreign Large-size Research Facility Application Supporting project, the Global Science Experimental Data Hub Center of the Korea Institute of Science and Technology Information and KREONET/GLORIAD;
the Polish Ministry of Science and Higher Education and 
the National Science Center;
the Ministry of Science and Higher Education of the Russian Federation, Agreement No. 14.W03.31.0026, % from 15.02.2018
and the HSE University Basic Research Program, Moscow; % from 15.04.2021
University of Tabuk research Grants No.
S-1440-0321, No. S-0256-1438, and No. S-0280-1439 (Saudi Arabia);
the Slovenian Research Agency Grants No. J1-9124 and No. P1-0135;
Ikerbasque, Basque Foundation for Science, and the State Agency for Research
of the Spanish Ministry of Science and Innovation through Grant No. PID2022-136510NB-C33 (Spain);
the Swiss National Science Foundation; 
the Ministry of Education and the National Science and Technology Council of Taiwan;
and the U.S. Department of Energy and the National Science Foundation.
These acknowledgements are not to be interpreted as an endorsement of any
statement made by any of our institutes, funding agencies, governments, or
their representatives.
We thank the KEKB group for the excellent operation of the
accelerator; the KEK cryogenics group for the efficient
operation of the solenoid; and the KEK computer group and the Pacific Northwest National
Laboratory (PNNL) Environmental Molecular Sciences Laboratory (EMSL)
computing group for strong computing support; and the National
Institute of Informatics, and Science Information NETwork 6 (SINET6) for
valuable network support.

\bibliographystyle{apsrev4-1}
\bibliography{note.bib}

%\begin{thebibliography}{99}

%\bibitem{pdg-2022}
%R. L.~Workman {\it et al.} (Particle Data Group), Prog. Theor. Exp. Phys. {\bf 2022}, Issue 8, 083C01 (2022).

%\bibitem{Minkowski:1977sc} 
%P.~Minkowski, Phys. Lett. B {\bf 67}, 421-428 (1977).

%\bibitem{Yanagida:1979as} 
%T.~Yanagida, Conf. Proc. C {\bf 7902131}, 95-99 (1979).

%\bibitem{geant3} R. Brun, F. Bruyant, M. Maire, A.C. McPherson, and P. Zanarini, CERN Report No. CERN-DD-EE-84-1, 2008.

%\bibitem{Sanda}
%A.~B.~Carter and A.~I.~Sanda, Phys. Rev. Lett. {\bf 45}, 952 (1980); 
%A.~B.~Carter and A.~I.~Sanda, Phys. Rev.  {\bf D23}, 1567 (1981); 
%I.~I.~Bigi and A.~I.~Sanda, Nucl. Phys. {\bf 193}, 85 (1981).

%\bibitem{CPVrev}
%A general review of the formalism is given in
%I.I.~Bigi, V.A.~Khoze, N.G.~Uraltsev, and A.I.~Sanda, ``$CP$ Violation''
%page 175, ed. C.~Jarlskog, World Scientific, Singapore (1989). 

%\end{thebibliography}
%

\begin{figure}[htb]
\begin{center}   
\includegraphics[width=1\linewidth]{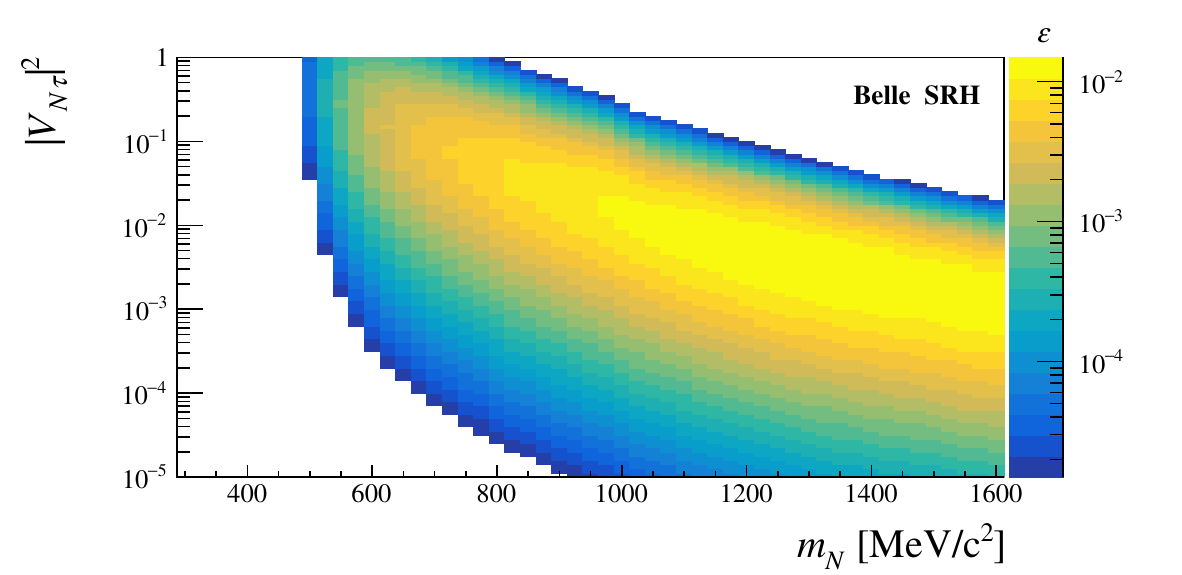}\\
\includegraphics[width=1\linewidth]{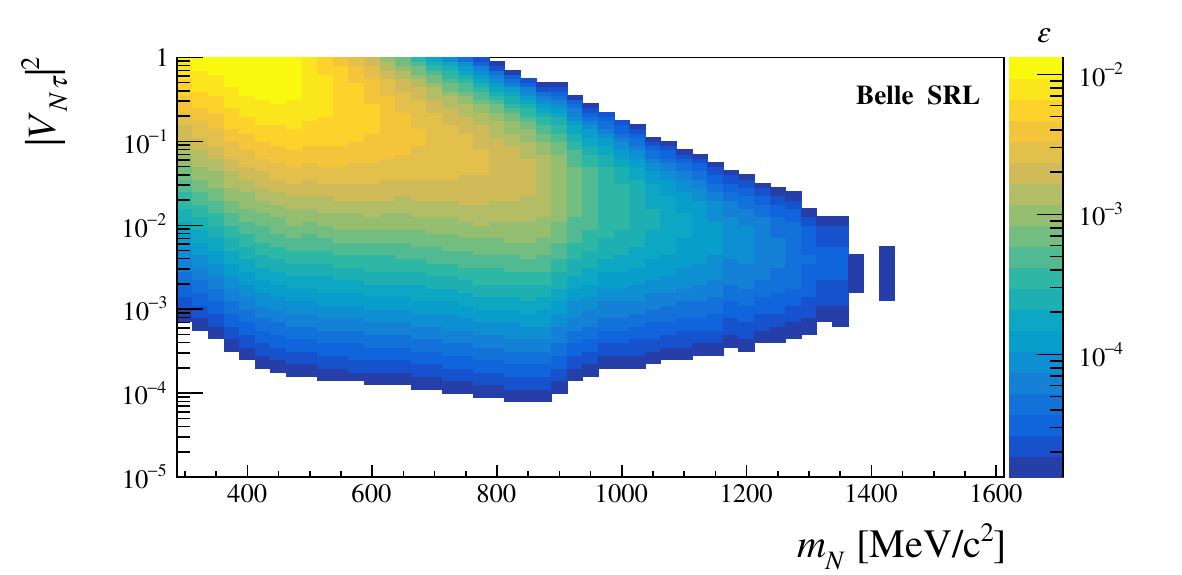}\\
\end{center}
\caption{The total signal efficiencies in (top) SRH and (bottom) SRL as a function of $\Vtausq$ and $m_N$.
}
\label{fig:effs} 
\end{figure}

\begin{figure}[!htbp]
\begin{center}
\includegraphics[trim={2cm 7.2cm 1cm 6cm},clip, width=\linewidth]{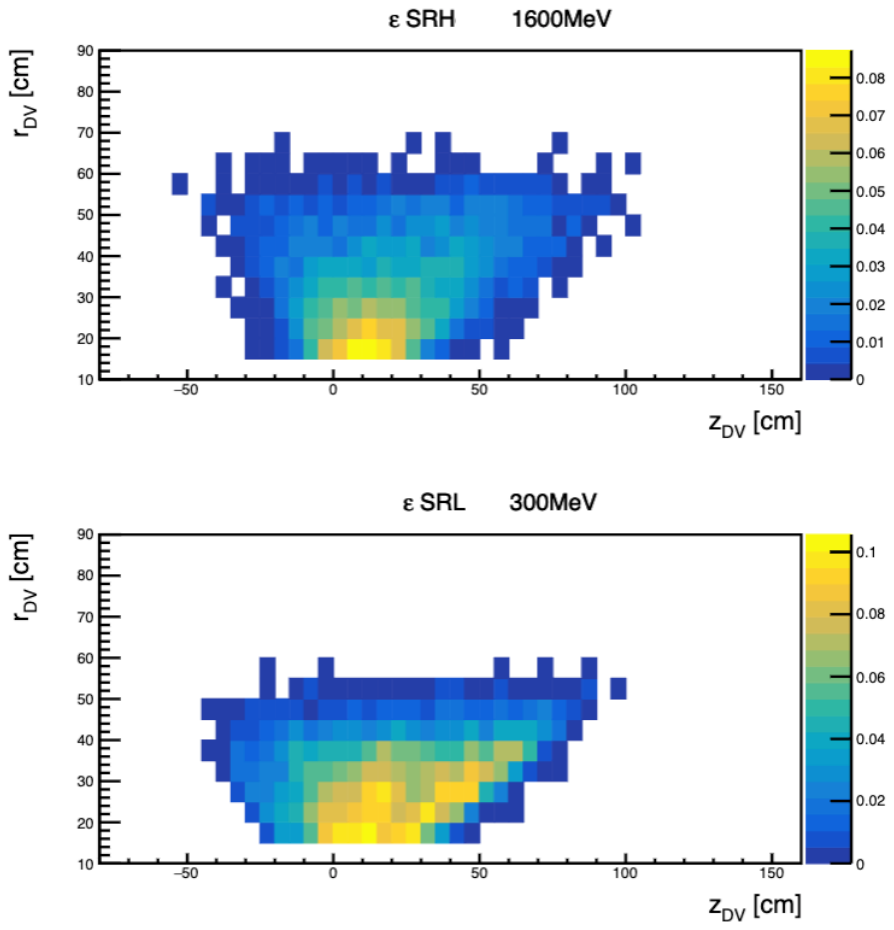}
\end{center}
\caption{Final reconstruction efficiency for signal events as a function of the radial and longitudinal positions of the DV for (top) $m_N=1600$~\mevcc events in the SRH and (bottom) $m_N=300$~\mevcc events in the SRL.} 
\label{fig:effmaps} 
\end{figure}

\appendix
\bigskip
{\sl Appendix: Efficiency. }Figure~\ref{fig:effs} shows the total efficiency for signal selection as a function of the mixing parameter $\Vtausq$ and the HNL mass $m_N$.
Examples of the efficiency maps are shown for $m_N=300$, $1600$~\mevcc in Fig.~\ref{fig:effmaps}.
The efficiency for the various off-line selection requirements (cut flow) is shown in Table~\ref{tab:cutflow-SR} for selected signal samples. 

\begin{table}[!htbp]
\begin{tabular}{|lllll|}
\hline\hline
& HNL mass (\gevcc)   & 0.3 & 1.0 & 1.6 \\
& Generated $c\tau$ & 15 & 22.5 & 30 \\
& {Selection criteria} & \multicolumn{3}{c|}{Efficiency (\%)} \\  \hline
& Online                                              & 3.7             & 4.0              & 3.9              \\ 
& 4 tracks                & 3.5             & 3.6              & 3.3              \\ 
&{$\rdv>15$~cm}                 & 2.1             & 2.2              & 1.8              \\ 
& DV $P(\chi^{2})>10^{-5}$ & 2.0             & 2.0              & 1.6              \\ 
&{$\cos\theta_{\mu^+\mu^-}>0.5$}  & 2.0             & 2.0              & 1.6              \\ 
&Lepton veto for pion    & 1.8             & 1.8              & 1.5              \\ 
& 20 CDC hits                                      & 1.6             & 1.7              & 1.3              \\ 
& Tight $\mu$-ID for one muon & 1.5             & 1.7              & 1.3              \\ 
& $\mpipi \notin [0.42,0.52]$~GeV     & 1.5             & 1.4              & 1.2              \\ 
&{$\sum E_\gamma<1$~GeV}           & 1.5             & 1.4              & 1.2              \\ 
& No ${\pi^0}$ candidates               & 1.4             & 1.4              & 1.1              \\ 
& $S<0.4$~GeV                  & 1.4             & 1.3              & 1.1              \\ 
& Tight $\mu$-ID for both muons       & 1.3             & 1.2              & 1.0              \\
\hline
SRH: & {{$\mpipi>0.52$}}~\gevcc      & 0               & 1.1              & 1.0              \\ 
& {{$\mpipi<0.42$}}~\gevcc      & 1.3             & 0.14             & 0                \\ 
SRL: & $m_+ <0.9$ or $m_- < 0.6~\gevcc$                                        & 1.3             & 0.021            & 0                \\ \hline\hline
\end{tabular}
\caption{Cut flow table showing the selection efficiency at different stages of the off-line selection for three signal samples, with masses of 300, 1000, and 1600~\mevcc, generated with $c\tau$ values of 15, 22.5, and 30~cm, respectively.
The first row gives the efficiency for the online selection, which includes the requirement $\rdv>5$~cm, loose muon identification for the muon candidates, and that the tag-$\tau$ track is in the opposite hemisphere to the three signal-$\tau$ tracks.  
In the leftmost column, the final selections for SRH and SRL are labeled.
} 
\label{tab:cutflow-SR}
\end{table}

%%%%%%%%%%%%%%%%%%%%%
{\sl Origin of DV tracks in MC events in the VR$\pi\pi$ regions.} Table~\ref{tab:VRpipi-types} gives details of the origin of the DV tracks in $e^+e^-\to\tau^+\tau^-$ and $e^+e^-\to q\bar q$ MC events. 

\begin{table}[!htb]
\centering
\begin{tabular}{|l|r|r|r|r|}
\hline\hline
Region  &\multicolumn{2}{c|}{VRH$\pi\pi$} & \multicolumn{2}{c|}{VRL$\pi\pi$} \\ 
\hline
Event sample & $\tau\tau$ & $q\bar q$    & $\tau\tau$ & $q\bar q$ \\
Total no. of events & 660 & 281 & 182 & 423 \\
\hline
  & \multicolumn{4}{c|}{Event type (\%)} \\  \hline
$K_s\to\pi\pi$                 & 86.7 & 12.8  & 0    & 0.5  \\
$\Lambda \to p\pi$             & 0    & 81.5  & 0    & 95.5 \\
$K_L$ decay                    & 0.5  & 0.4   & 37.4 & 2.1 \\
Hadronic material interaction  & 3.9  & 1.2  & 4.9   & 1.4 \\
$K/\pi$ decay in flight        & 7.7  & 1.8  & 1.1   & 0   \\
2 Prompt tracks                & 0.9  & 1.4  & 0.5   & 0   \\
$\gamma$ conversion            & 0.3  & 0    & 53.8  & 0.5 \\
Insufficient truth information & 0    & 0    & 2.2   & 0   \\
\hline\hline
\end{tabular}
\caption{Origin of the tracks forming the DV in the VRH$\pi\pi$ and VRL$\pi\pi$ validation regions in the MC sample, whose integrated luminosity is 5 times that of the data.
The different categories are mutually exclusive. 
In the ``hadronic material interaction'' and $K/\pi$ decay in flight categories either one or both tracks may be of this origin.}
\label{tab:VRpipi-types}
\end{table}

\begin{figure}[htb]
\begin{center}   
\includegraphics[width=1\linewidth]{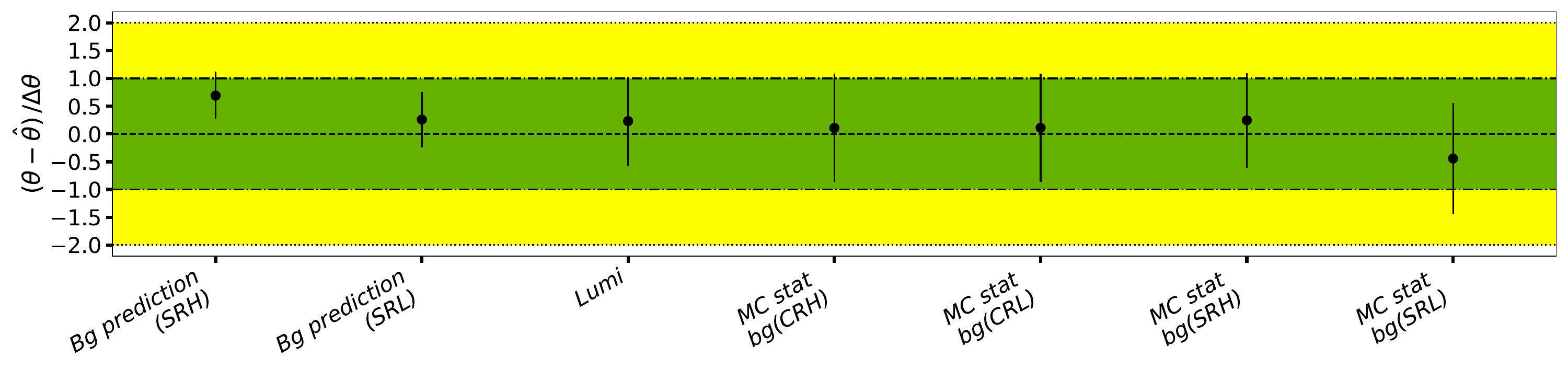}
\end{center}
\caption{The pull values of the nuisance parameters for the fit to data with $\mu$ fixed to 0.
The nuisance parameters correspond to the systematic uncertainties arising from (shown from left to right) the background prediction in SRH and SRL; the product of the integrated luminosity $\cal L$, $e^+e^-\to \tau^+\tau^-$ cross section $\sigma$, and the branching fractions $\BR_\tau$ and $\BR_N$ for the signal $\tau$ and HNL decays; and the background sample sizes in the CRH, CRL, SRH, SRL.
}
\label{fig:pulls} 
\end{figure}

{\sl Estimation of Nuisance Parameter pulls.} Figure~\ref{fig:pulls}  shows the pulls for each of the nuisance parameters when the expected signal yield is zero.

%%%%%%%%%%%%%%%%%%%%%%%%
{\sl Calculation of the HNL mass values $m_\pm$.}
Consider the decay chain $\tau\to Nx$, $N\to \nu_\tau y$, where the kinematics of the $x$ and $y$ systems are measured.
We start with 4-momentum conservation in the $\tau$ decay, 
$p_{\tau} = p_{N} + p_{x}$.
Squaring this relates the HNL energy $E_N$, momentum $|\vec p_{N}|$, and mass $m_N$ to those of the $x$ and to the $\tau$ mass, 
\begin{eqnarray}
m^2_{\tau} &=& m^2_{N} + m^2_{x} + 2E_{N}E_{x} - 2|\vec p_{N}||\vec q_{x}|~, 
\label{eq:4-momentum_tauV}
\end{eqnarray}
where \begin{eqnarray}
|\vec q_{x}| \equiv |\vec p_{x}|\cos\theta_{Nx}.
\end{eqnarray}
Here $\theta_{Nx}$ is the angle between the momentum directions of the $x$ and the HNL, the latter given by the position of the DV relative to the IP, where the small flight distance of the $\tau$ can be neglected.
Similarly, 4-momentum conservation in the HNL decay yields
\begin{eqnarray}
 0 &=& m^2_{N} + m^2_{y} - 2E_{N}E_{y} + 2|\vec p_{N}||\vec q_{y}| \ , 
 \label{eq:4-momentum_hnlV}
\end{eqnarray}
where \begin{eqnarray}
|\vec q_{y}| \equiv |\vec p_{y}|\cos\theta_{Ny}
\end{eqnarray}
and $\theta_{Ny}$ is the angle between the momenta of the $y$ and the HNL.
Equations~(\ref{eq:4-momentum_tauV}) and (\ref{eq:4-momentum_hnlV}) give $E_N$ in terms of $|\vec p_{N}|$:
\begin{eqnarray}
E_N &=&  A +  B|\vec p_{N}|~,  \label{eq:energy_hnlV}
\end{eqnarray}
where 
\begin{eqnarray}
A = \frac{m^2_{\tau} + m^2_{y} - m^2_{x}}{2(E_{x}+E_{y})}, ~
B = \frac{(|\vec q_{y}| + |\vec q_{x}|)}{(E_{x}+E_{y})}
\end{eqnarray} 
are known.
The HNL mass is then
\begin{eqnarray}
m^2_{N} = (A + B|\vec p_{N}|)^2  - |\vec p_N|^2~ 
\label{eq:m2_N_sol1}
\end{eqnarray}
and is also given by inserting  Eq.~(\ref{eq:energy_hnlV}) in (\ref{eq:4-momentum_tauV}),
\begin{eqnarray}
m^2_{N} &=& C + D|\vec p_{N}|~, 
\label{eq:m2_N_sol2}
\end{eqnarray}
where
\begin{eqnarray}
C &=& \frac{E_y(m^2_{\tau} - m^2_{x}) - {E_x} m^2_{y}}{({E_x} +{E_y})}, \nonumber\\
D &=& \frac{2(E_y|\vec q_{x}| - {E_x}|\vec q_{y}|)}{({E_x}+{E_y})}
\end{eqnarray} 
are known.
Comparing Eqs.~(\ref{eq:m2_N_sol1}) and~(\ref{eq:m2_N_sol2}) gives the quadratic equation
\begin{eqnarray}
 (B^2 - 1)|\vec p_N|^2 + (2AB - D)|\vec p_N| + (A^2 - C) &=& 0~~\; 
\end{eqnarray}
with the solution 
\begin{eqnarray}
 |\vec p_N| &=& \frac{-(2AB - D) \pm \sqrt{S}}{2(B^{2}-1)}~, 
 \label{eq:p_N_sol}
\end{eqnarray}
where $S=(2AB-D)^{2} - 4(B^{2}-1)(A^{2}-C)$.
Inserting Eq.~(\ref{eq:p_N_sol}) into Eq.~(\ref{eq:m2_N_sol2}) gives the two solutions for $m_N$.

{\sl HNL lifetime and branching fractions.}
Table~\ref{tab:HNLpars} lists the values used for the HNL lifetime, its production branching fraction in $\tau$ decay, and the branching fraction for its decay into $\mu\mu\nu$.

\begin{table}
\caption{The HNL lifetime times the speed of light and the branching fractions for $\tau\to \pi N$  and $N\to \mu\mu\nu$ as functions of the HNL mass, normalized by the mixing coefficient $\Vtausq$ where relevant.}
\label{tab:HNLpars} 
\begin{tabular}{|llll|}
\hline\hline
$m_N$\;(GeV) & $\dfrac{c\tau_N \; \textrm{(mm)}}{\Vtausq}$ & $\dfrac{{\cal B}(\tau\to \pi N)}{\Vtausq}$ & ${\cal B}(N\to \mu\mu\nu)$ \\
\hline
$0.225 $&$	1.64\times 10^3 $&$	1.07\times 10^{-1} $&$	3.78\times 10^{-6} $\\
$0.25 $&$	9.68\times 10^2 $&$	1.06\times 10^{-1} $&$	9.47\times 10^{-5} $\\
$0.275 $&$	6.28\times 10^2 $&$	1.05\times 10^{-1} $&$	3.81\times 10^{-4} $\\
$0.3 $&$	4.32\times 10^2 $&$	1.03\times 10^{-1} $&$	8.95\times 10^{-4} $\\
$0.325 $&$	3.12\times 10^2 $&$	1.02\times 10^{-1} $&$	1.64\times 10^{-3} $\\
$0.35 $&$	2.33\times 10^2 $&$	9.98\times 10^{-2} $&$	2.59\times 10^{-3} $\\
$0.375 $&$	1.78\times 10^2 $&$	9.80\times 10^{-2} $&$	3.71\times 10^{-3} $\\
$0.4 $&$	1.39\times 10^2 $&$	9.61\times 10^{-2} $&$	5.00\times 10^{-3} $\\
$0.425 $&$	1.10\times 10^2 $&$	9.41\times 10^{-2} $&$	6.43\times 10^{-3} $\\
$0.45 $&$	8.88\times 10^1 $&$	9.20\times 10^{-2} $&$	7.94\times 10^{-3} $\\
$0.475 $&$	7.23\times 10^1 $&$	8.98\times 10^{-2} $&$	9.56\times 10^{-3} $\\
$0.5 $&$	5.97\times 10^1 $&$	8.76\times 10^{-2} $&$	1.13\times 10^{-2} $\\
$0.525 $&$	4.95\times 10^1 $&$	8.53\times 10^{-2} $&$	1.30\times 10^{-2} $\\
$0.55 $&$	4.16\times 10^1 $&$	8.29\times 10^{-2} $&$	1.48\times 10^{-2} $\\
$0.575 $&$	3.50\times 10^1 $&$	8.04\times 10^{-2} $&$	1.66\times 10^{-2} $\\
$0.6 $&$	2.97\times 10^1 $&$	7.79\times 10^{-2} $&$	1.84\times 10^{-2} $\\
$0.625 $&$	2.52\times 10^1 $&$	7.53\times 10^{-2} $&$	2.01\times 10^{-2} $\\
$0.65 $&$	2.16\times 10^1 $&$	7.27\times 10^{-2} $&$	2.18\times 10^{-2} $\\
$0.675 $&$	1.85\times 10^1 $&$	7.00\times 10^{-2} $&$	2.34\times 10^{-2} $\\
$0.7 $&$	1.59\times 10^1 $&$	6.73\times 10^{-2} $&$	2.50\times 10^{-2} $\\
$0.725 $&$	1.38\times 10^1 $&$	6.46\times 10^{-2} $&$	2.66\times 10^{-2} $\\
$0.75 $&$	1.20\times 10^1 $&$	6.19\times 10^{-2} $&$	2.81\times 10^{-2} $\\
$0.775 $&$	1.05\times 10^1 $&$	5.92\times 10^{-2} $&$	2.97\times 10^{-2} $\\
$0.8 $&$	9.10 $&$	5.64\times 10^{-2} $&$	3.09\times 10^{-2} $\\
$0.825 $&$	7.88 $&$	5.37\times 10^{-2} $&$	3.19\times 10^{-2} $\\
$0.85 $&$	6.85 $&$	5.09\times 10^{-2} $&$	3.28\times 10^{-2} $\\
$0.875 $&$	5.99 $&$	4.82\times 10^{-2} $&$	3.37\times 10^{-2} $\\
$0.9 $&$	5.24 $&$	4.55\times 10^{-2} $&$	3.45\times 10^{-2} $\\
$0.925 $&$	4.59 $&$	4.29\times 10^{-2} $&$	3.51\times 10^{-2} $\\
$0.95 $&$	4.04 $&$	4.02\times 10^{-2} $&$	3.58\times 10^{-2} $\\
$0.975 $&$	3.56 $&$	3.76\times 10^{-2} $&$	3.64\times 10^{-2} $\\
$1 $&$	        3.01 $&$	3.51\times 10^{-2} $&$	3.53\times 10^{-2} $\\
$1.025 $&$	2.46 $&$	3.26\times 10^{-2} $&$	3.29\times 10^{-2} $\\
$1.05 $&$	2.18 $&$	3.02\times 10^{-2} $&$	3.33\times 10^{-2} $\\
$1.075 $&$	1.94 $&$	2.78\times 10^{-2} $&$	3.36\times 10^{-2} $\\
$1.1 $&$	1.73 $&$	2.56\times 10^{-2} $&$	3.40\times 10^{-2} $\\
$1.125 $&$	1.55 $&$	2.33\times 10^{-2} $&$	3.43\times 10^{-2} $\\
$1.15 $&$	1.39 $&$	2.12\times 10^{-2} $&$	3.46\times 10^{-2} $\\
$1.175 $&$	1.25 $&$	1.92\times 10^{-2} $&$	3.48\times 10^{-2} $\\
$1.2 $&$	1.13 $&$	1.72\times 10^{-2} $&$	3.51\times 10^{-2} $\\
$1.225 $&$	1.02 $&$	1.54\times 10^{-2} $&$	3.53\times 10^{-2} $\\
$1.25 $&$	9.21\times 10^{-1} $&$	1.36\times 10^{-2} $&$	3.56\times 10^{-2} $\\
$1.275 $&$	8.35\times 10^{-1} $&$	1.20\times 10^{-2} $&$	3.58\times 10^{-2} $\\
$1.3 $&$	7.59\times 10^{-1} $&$	1.05\times 10^{-2} $&$	3.60\times 10^{-2} $\\
$1.325 $&$	6.90\times 10^{-1} $&$	9.02\times 10^{-3} $&$	3.62\times 10^{-2} $\\
$1.35 $&$	6.29\times 10^{-1} $&$	7.69\times 10^{-3} $&$	3.64\times 10^{-2} $\\
$1.375 $&$	5.74\times 10^{-1} $&$	6.48\times 10^{-3} $&$	3.66\times 10^{-2} $\\
$1.4 $&$	5.25\times 10^{-1} $&$	5.37\times 10^{-3} $&$	3.68\times 10^{-2} $\\
$1.425 $&$	4.81\times 10^{-1} $&$	4.37\times 10^{-3} $&$	3.69\times 10^{-2} $\\
$1.45 $&$	4.41\times 10^{-1} $&$	3.49\times 10^{-3} $&$	3.71\times 10^{-2} $\\
$1.475 $&$	4.05\times 10^{-1} $&$	2.71\times 10^{-3} $&$	3.72\times 10^{-2} $\\
$1.5 $&$	3.73\times 10^{-1} $&$	2.04\times 10^{-3} $&$	3.74\times 10^{-2} $\\
$1.525 $&$	3.44\times 10^{-1} $&$	1.47\times 10^{-3} $&$	3.76\times 10^{-2} $\\
$1.55 $&$	3.17\times 10^{-1} $&$	9.95\times 10^{-4} $&$	3.77\times 10^{-2} $\\
$1.575 $&$	2.93\times 10^{-1} $&$	6.17\times 10^{-4} $&$	3.78\times 10^{-2} $\\
$1.6 $&$	2.71\times 10^{-1} $&$	3.27\times 10^{-4} $&$	3.80\times 10^{-2} $\\
\hline\hline
\end{tabular}    
\end{table}

\end{document}